\documentclass[a4paper,12pt]{article}
\usepackage{amssymb}
\usepackage[dvips]{epsfig}
\newcommand{\be}{\begin{equation}}
\newcommand{\ee}{\end{equation}}
\newcommand{\bea}{\begin{eqnarray}}
\newcommand{\eea}{\end{eqnarray}}
\newcommand{\bean}{\begin{eqnarray*}}
\newcommand{\eean}{\end{eqnarray*}}
\newcounter{saveeqn}

\newcounter{doeqn}

\newcommand{\cl}[1]{\mathcal{#1}}
\newcommand{\text}[1]{\mbox{#1}}
\newcommand{\dts}[2]{#1_{{\scriptsize \mbox{#2}}}}

\newcommand{\Tr}{\mbox{Tr\,}}
\newcommand{\tr}[1]{\mbox{Tr}\left(#1\right)}

\newcommand{\dt}{\mbox{d}}

\newcommand{\ltx}{L_{{\scriptsize \mbox{x}}}}
\newcommand{\lty}{L_{{\scriptsize \mbox{y}}}}
\newcommand{\ltz}{L_{{\scriptsize \mbox{z}}}}
\newcommand{\ltzs}{L_{{\tiny \mbox{z}}}}
\newcommand{\lts}{L_{{\scriptsize \mbox{s}}}}
\newcommand{\x}{\mbox{x}}
\newcommand{\y}{\mbox{y}}
\newcommand{\z}{\mbox{z}}
\newcommand{\mbs}{\mbox{s}}
\newcommand{\hf}[1]{#1_{x}}
\newcommand{\hfup}[2]{#1_{x+\hat{#2}}}

\newcommand{\gf}[2]{#1_{x,#2}}
\newcommand{\gfup}[3]{#1_{x+\hat{#2},#3}}

\newcommand{\lkop}[3]{#1_{#2;#3}}
\newcommand{\half}{{\textstyle{\frac{1}{2}}}}

\title{
{
\vspace{-3cm} \normalsize \hfill
\parbox{35mm}{DESY 96-066\\MS-TPI-96-7\\hep-lat/9605009}
}\\[20mm]
\textbf{
On the scaling of the electroweak\\
interface tension at finite temperature}
}
\author{
J.~Hein
\\[0.5mm]
\normalsize\textsl{
Deutsches Elektronen-Synchrotron DESY,}\\[-1.0mm]
\normalsize\textsl{
Notkestr.~85, D-22603 Hamburg, Germany}\\[-1.0mm]
\\[-3.0mm]
J.~Heitger
\\[0.5mm]
\normalsize\textsl{
Institut f\"ur Theoretische Physik I, Universit\"at M\"unster,}\\[-1.0mm]
\normalsize\textsl{
Wilhelm-Klemm-Str.~9, D-48149 M\"unster, Germany}\\[-1.0mm]
}
\date{}
\begin{document}
\maketitle
\begin{abstract} \noindent
%
%%%%%%%%%%%%%%%%%%%%%%%%%%%%%%%%%%%%%%%%%%%%%%%%%%%%%%%%%%%%%%%%%%%%%%%%%%%
We determine the interface tension of the finite-temperature electroweak
phase transition in a numerical investigation of the SU(2)--Higgs model on
a four-dimensional lattice with temporal extension $L_t=3$.
In this simulation the chosen parameters correspond to a Higgs boson mass
of about 16 GeV.
As a result the interface tension shows only small scaling violations in
comparison with previous studies for $L_t=2$ lattices.
We also report on some experiences with autocorrelations in the applied
Monte Carlo simulations of two-phase systems.
%%%%%%%%%%%%%%%%%%%%%%%%%%%%%%%%%%%%%%%%%%%%%%%%%%%%%%%%%%%%%%%%%%%%%%%%%%%%
%
\end{abstract}
%
% \newpage
%
%%%%%%%%%%%%%%%%%%%%%%%%%%%%%%%%%%%%%%%%%%%%%%%%%%%%%%%%%%%%%%%%%%%%%%%%%%%%
%
\section{Introduction}
According to the ideas of cosmology and elementary particle physics the
electroweak phase transition takes place with decreasing temperature
between a \emph{symmetric phase} and a phase with broken symmetry, the
\emph{Higgs phase}.
An order parameter of this transition is the vacuum expectation value
of the scalar field, which has a non-zero value in the broken phase and
vanishes in the high-temperature, symmetry-restored phase \cite{K72KL7276}.

The main interest in the electroweak phase transition in the early universe
emerges from the question, whether it alone can provide a mechanism for the
observed baryon asymmetry within the minimal Standard Model
\cite{KRS85S87}.
A necessary condition for this scenario, essentially depending on the
Higgs boson mass value $m_H$, is a transition of strong enough first order
type, whereas for a weak transition every asymmetry generated at the 
phase transition would be washed out in the Higgs phase, because at 
temperatures $T$ larger than the vector boson mass $m_W$ the baryon and 
lepton number violation in the Standard Model is enhanced.

After the neglection of the U(1)--gauge field and the fermionic sector one is
left with the SU(2)--Higgs model containing all important ingredients for
these phenomena.
The common theoretical tool for its examination is resummed perturbation
theory \cite{AE93,BFHW94,FH94BFH9495}, which works well for low and
intermediate Higgs masses ($m_H<50$ GeV).
Besides the known problems caused by the infrared divergencies in the
symmetric phase, its validity is not ensured any longer for higher values of
$m_H$, where the strength of the electroweak phase transition is expected
to decrease rapidly.
This motivates a nonperturbative treatment on space-time lattices
in four dimensions as well as in reduced three-dimensional models, see 
e.g.~refs.~\mbox{
\cite{BIKS9293IS95,KNP95,KRS93FKRS9495KLRS95,IKPS95GIKPS95}
}
and ref.~\cite{J95} for a more general review of the whole subject.

The present work is a completion of numerical simulations of the
four-dimensional SU(2)--Higgs model on the lattice in
\cite{CFHJJM9495,FHJJM95}, which focussed on Higgs boson masses below 50 GeV.
Since this mass range is ruled out by the actual experimental bound of 
$m_H\gtrsim 65$ GeV, their main goal is to establish suitable
methods for extracting physical quantities from the lattice investigations,
and to hint at systematic errors in resummed perturbation theory or 
dimensional reduction approaches.

In this publication we deal with the interface tension $\sigma$.
This quantity plays a prominent r\^ole in the course of the electroweak phase
transition, because its magnitude is a measure for the strength of this
transition.
The nucleation rate of the Higgs phase in the symmetric phase is given in 
the thin-wall approximation by \cite{L83}
\be
\dts{\Gamma}{nucl}=\Gamma_0\exp\left(-\frac{16\pi}{3}
                   \frac{\sigma^3}{(\Delta\epsilon)^2 T_c}\,
                   \eta^{-2}\right)\,.
\ee
Here $T_c$ denotes the critical temperature, $\Delta\epsilon$ the latent
heat, and $\eta\equiv\frac{T_c-T}{T_c}$ is the so-called supercooling
parameter.
The prefactor $\Gamma_0$ can be approximated by $T^4$ \cite{BFHW94}.
If $\sigma^3/(\Delta\epsilon)^2 T_c$ is large, signalling a strong first
order phase transition, a substantial supercooling has to be expected, which
would lead to an additional suppression of the sphaleron rate.

In the following we use a low value of the Higgs boson mass, where the phase
transition is quite strong.
For the determination of $\sigma$ the two-coupling method in the scalar
hopping parameter is employed.
Compared to transfer-matrix techniques and histogram methods, it gives an
optimal ratio between desired accuracy and required CPU-time for the
SU(2)--Higgs model \cite{CFHJJM9495,FHJJM95,CFHH95}.
\section{Lattice simulation}
The lattice action of the SU(2)--Higgs model is conventionally \cite{MM94}
parametrized as
\bea
S[U,\varphi]
& = &\beta\dts{\sum}{p}\Big(1-\half\Tr\dts{U}{p}\Big)
     \nonumber\\
&   &+\sum_{x\in\Omega}\Bigg\{\half\tr{\hf{\varphi}^+\hf{\varphi}}
     +\lambda\Big[\half\tr{\hf{\varphi}^+\hf{\varphi}}
     -1\Big]^2-\kappa\sum_{\mu=1}^{4}\tr{\hfup{\varphi}{\mu}^+
                 \gf{U}{\mu}\hf{\varphi}}\Bigg\}
\label{lattact}
\eea
in terms of the SU(2)--link variables $\gf{U}{\mu}$ and the site variables
$\hf{\varphi}$, which are complex $2\otimes 2$ matrices, representing the
gauge and scalar degrees of freedom, respectively.
\mbox{
$\dts{U}{p}=\gf{U}{\mu}\gfup{U}{\mu}{\nu}\gfup{U}{\nu}{\mu}^+\gf{U}{\nu}^+$
}
is an elementary plaquette, and we often decompose the Higgs field
as $\hf{\varphi}=\hf{\rho}\hf{\alpha}$ with $\hf{\rho}\in\Bbb{R}^{>0}$ and
$\hf{\alpha}\in\mbox{SU(2)}$.
If not stated otherwise, the lattice constant $a$ is assumed to be $a=1$,
and the lattice volume is denoted by $\Omega$.
The identifications $g^2=4/\beta$, $m_0^2=(1-2\lambda)/\kappa-8$
and $\lambda_0=\lambda/4\kappa^2$ relate the lattice parameters $\beta$,
$\kappa$ and $\lambda$ to the bare gauge coupling, scalar mass and quartic
coupling of the corresponding continuum theory.

A suitable observable for the interface tension is the density of the
$\varphi$--link operator $\lkop{L}{\varphi}{x\mu}$, which is conjugate
--- in the thermodynamic sense --- to the hopping parameter $\kappa$ and
is itself an order parameter of the phase transition:
\be
L_{\varphi}\equiv\frac{1}{4\Omega}\sum_{x\in\Omega}\sum_{\mu=1}^4
                 \lkop{L}{\varphi}{x\mu}\,,\quad
\lkop{L}{\varphi}{x\mu}
           \equiv\half\tr{\hfup{\varphi}{\mu}^+\gf{U}{\mu}\hf{\varphi}}\,.
\label{lphidef}
\ee

When simulating finite-temperature field theory, one utilizes lattices with
spacelike extensions much larger than the temporal extension: 
$\lts\gg L_t$, $\mbs\in\{\x,\y,\z\}$.
The physical temperature is given by the timelike lattice extension via
$T=1/aL_t$, and the approach to the scaling region of the model in the
continuum limit is realized as $a\rightarrow 0$ with $T$ fixed, i.e.~as
$L_t\rightarrow\infty$.
This limit goes along the \emph{lines of constant physics}, on which
renormalized couplings and masses are held fixed and only the lattice 
constant $a$ is varying.

In this spirit we extend the measurement of $\sigma$ at $a^{-1}=2T_c$ in
ref.~\cite{FHJJM95} to $a^{-1}=3T_c$ with the purpose of gaining control
over possible lattice artifacts.
The parameters in the former analysis were $\beta=8.0$ and $\lambda=0.0001$,
leading to a renormalized gauge coupling of $g_R^2\simeq 0.56$ and a Higgs
mass of $m_H\simeq 16$ GeV.
The physical mass scale is set by the vector boson mass value
$m_W=80$ GeV at $T=0$ in lattice units.
A two-coupling simulation on a lattice of size $2\times16^2\times128$
resulted in a phase transition point at $\kappa_c=0.12830(5)$ and a
finite-volume estimator for the interface tension of
$\hat{\sigma}/T_c^3=0.84(16)$.

When passing over to smaller lattice spacings, we have to scale all lattice
extensions accordingly in order to keep the physical volume constant.
In the case of \emph{low} Higgs mass this is possible with an acceptable
demand of computer resources.
With increasing $m_H$ the situation becomes worse, because the phase
transition weakens and thus one needs larger physical volumes to obtain
a stable two-phase situation.
Therefore we choose a lattice of size $3\times24^2\times192$ with the
changed parameters $\beta=8.15$ and $\lambda=0.00011$.
They have been obtained by an integration of the one-loop perturbative
renormalization group equations with the transition point of the $L_t=2$
lattice as a starting value \cite{FHJJM95}.
As shown in ref.~\cite{CFHJM96}, a sufficiently precise estimate for the
critical point in $\kappa_c$ is only available by numerical methods.

\begin{figure}[h]
\begin{center}
\epsfig{file=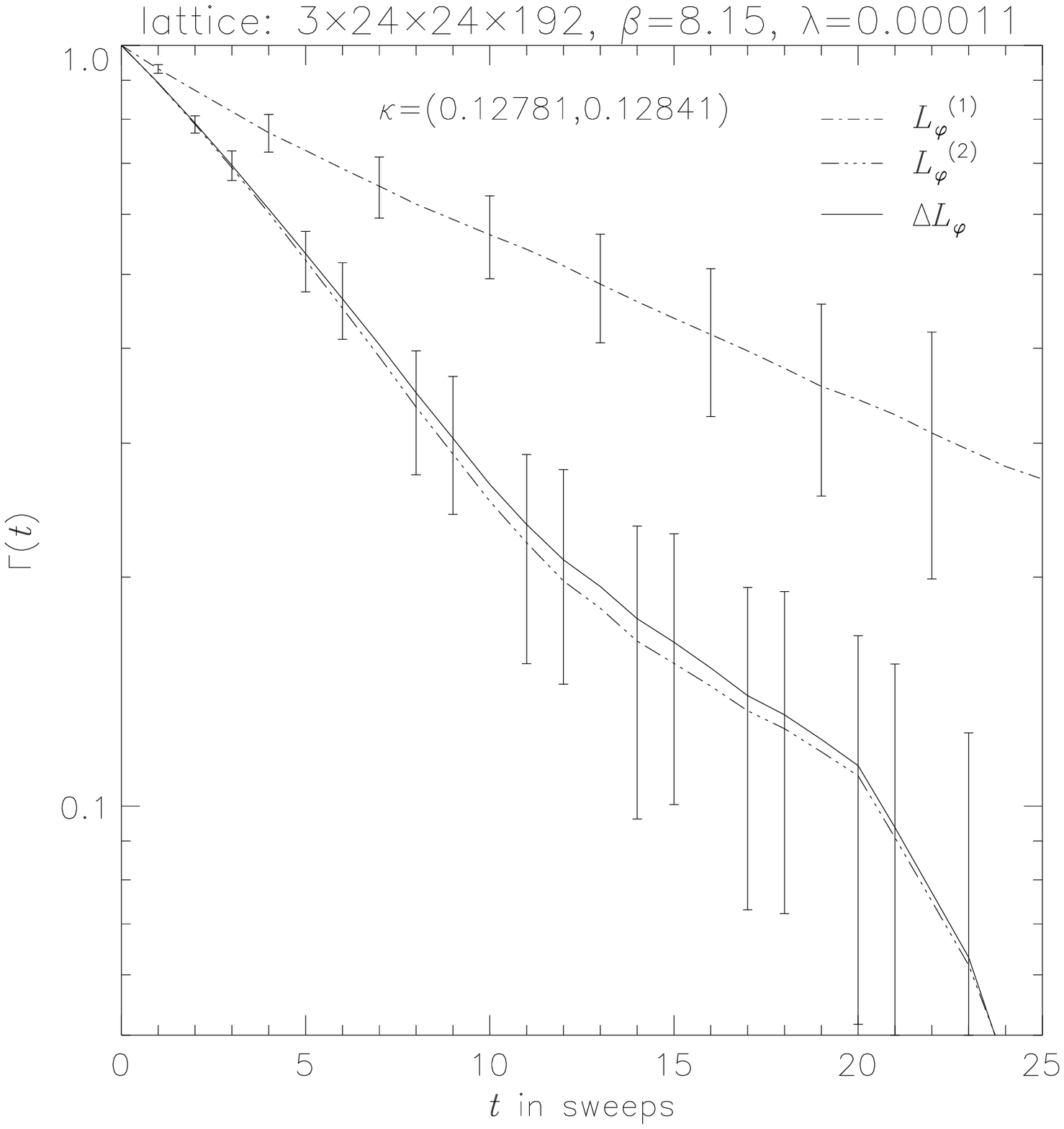,width=8cm}
\epsfig{file=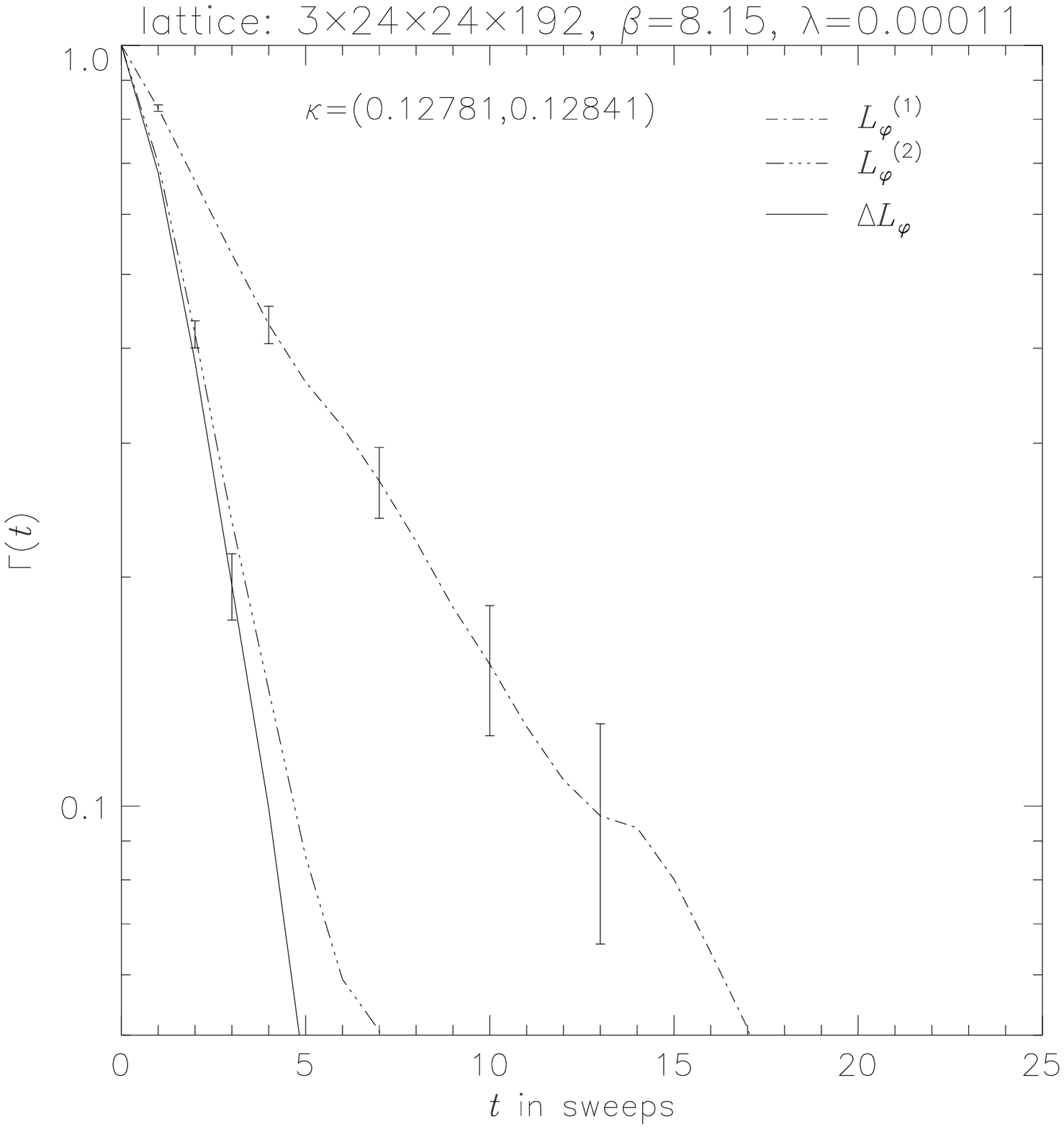,width=8cm}
\parbox{16cm}{
\caption{\label{AcfPlot} \sl Normalized autocorrelation functions for a
                             2--$\kappa$ simulation with
                             $\rho$--overrelaxation (left) and
                             $\varphi$--overrelaxation (right) in the
                             updating sequence.}
}
\end{center}
\end{figure}

In our Monte Carlo simulations we use an optimized combination of heatbath
and overrelaxation algorithms.
In particular, the introduction of the simultaneous overrelaxation of all 
four cartesian components of the Higgs field \cite{B95}, instead of the 
Higgs field length-overrelaxation as proposed in \cite{FJ94}, has 
substantially reduced the integrated autocorrelation time $\dts{\tau}{int}$.
In this way we gained factors of $3-10$ compared to the values previously
found in \cite{CFHH95}.
This $\dts{\tau}{int}$--behaviour is reflected in table~\ref{AcfTab} and
figure~\ref{AcfPlot} for the case of a two-phase simulation, where as
typical examples the autocorrelation functions $\Gamma(t)$ of the
$\varphi$--link operators (\ref{lphidef}) in both phases and their difference
$\Delta L_{\varphi}$ are considered.
The observation that this difference has a significantly lower
$\dts{\tau}{int}$ will become relevant in the next section.
%
%%% Beginn Tabelle %%%
\begin{table}[h]
\begin{center}
\begin{tabular}{|c|c|c|c|c|c||c|c|c|}
\hline
  \multicolumn{2}{|c|}{\rule[-3mm]{0mm}{8mm} heatbath }
& \multicolumn{4}{|c||}{\rule[-3mm]{0mm}{8mm} overrelaxation }
& \multicolumn{3}{|c|}{\rule[-3mm]{0mm}{8mm} $\dts{\tau}{int}$ in sweeps } \\
\hline
  $\gf{U}{\mu}$ & $\hf{\varphi}$ & $\gf{U}{\mu}$ & $\hf{\alpha}$
& $\hf{\rho}$ & $\hf{\varphi}$ & $L_{\varphi}^{(1)}$ & $L_{\varphi}^{(2)}$
& $\Delta L_{\varphi}$ \\
\hline\hline
  1 & 4 & 3 & 3 & 1 & - & 20(6) & 9.4(3.3) & 9.2(3.1) \\
  1 & 1 & 1 & - & - & 3 & 6.3(1.1) & 2.2(4) & 2.0(4) \\
\hline
\end{tabular}
\parbox{11cm}{
\caption{\label{AcfTab} \sl Autocorrelation times for a 2--$\kappa$
                            simulation. Each updating sweep consists of a
                            sequence of different algorithms as given by the
                            numbers in the left part of the table.}
}
\end{center}
\end{table}
%%% Ende Tabelle %%%
%
%%% Beginn Figur %%%
%%% Ende Figur %%%
%
\section{Two-coupling method}
The calculation of the phase transition point and the interface tension is
done by a modified version of the \emph{two-coupling method}
\cite{PR89HPRS9091}.
Since this method has been used in the SU(2)--Higgs model before, see
refs.~\cite{CFHJJM9495,FHJJM95,CFHH95}, we only sketch its main idea here.
One takes a periodic lattice with one elongated spacelike direction,
i.e.~$\ltz\gg\ltx=\lty\gg L_t$, and divides the corresponding lattice 
volume into two halves with different scalar hopping parameters
\be
\kappa=(\kappa_1,\kappa_2)
 \equiv(\kappa_1<\kappa_c\,\,\mbox{for}\,\,\z\le\ltz/2\,,\,
        \kappa_2>\kappa_c\,\,\mbox{for}\,\,\z>\ltz/2)\,.
\label{kapdef}
\ee
Hence the lower half is forced in the symmetric phase and the upper one
in the Higgs phase.
If in the foregoing inequality the transverse directions $\ltx$ and $\lty$
are sufficiently large, the system resides in a \emph{mixed-phases state}
and gives rise to an interface pair at the phase boundary perpendicular to
the $\z$--direction.

In a first step one initializes a two-phase situation by a simulation with
$\kappa$--values far away from the transition point, whose location can
roughly be determined by hysteresis runs.
Subsequently, the distance in $\kappa$ is more and more diminished, and
the smallest $\kappa$--interval, for which the system still resists
to turn over into one single phase, gives lower and upper bounds for the
critical $\kappa$.
On the $3\times24^2\times192$ lattice we obtain the estimate
\be
\kappa_c=0.128110(3)\,.
\label{kappares}
\ee

The interface tension is defined as the free energy $F$ per unit area of the
walls separating the two phases.
As derived in ref.~\cite{FHJJM95}, its lattice version
\be
a^3\sigma=\frac{1}{2\ltx\lty L_t}\,
          \Big\{F(\kappa_1,\kappa_2)-\half F(\kappa_1,\kappa_1)
                                    -\half F(\kappa_2,\kappa_2)\Big\}
\label{sigmadef}
\ee
is related to the expectation values $L^{(1)}_{\varphi}(\kappa_1,\kappa_2)$
and $L^{(2)}_{\varphi}(\kappa_1,\kappa_2)$ of $L_{\varphi}$ in each phase by
\be
a^3\sigma
=\frac{1}{2}\,\lim_{\kappa_2\searrow\kappa_c}\,
 \lim_{\kappa_1\nearrow\kappa_c}\left\{\left(\kappa_1-\kappa_2\right)
 \lim_{\ltzs\to\infty}\ltz\cdot\Delta L_\varphi(\kappa_1,\kappa_2)
 \right\}\,,
\label{sigmares}
\ee
where $\Delta L_{\varphi}(\kappa_1,\kappa_2)$ denotes their difference
$L^{(2)}_{\varphi}(\kappa_1,\kappa_2)-L^{(1)}_{\varphi}(\kappa_1,\kappa_2)$.
With the $(N+2)$--parametric Laurent ansatz
\be
L^{(i)}_{\varphi}(\kappa_1,\kappa_2)
=-\frac{c_i}{\kappa_i-\kappa_c}
 +\sum_{j=0}^{N}\gamma^{(j)}_i(\kappa_i-\kappa_c)^j
 +\cl{O}\left(|\kappa_i-\kappa_c|^{N+1}\right),\quad i=1,2\,,
\label{fitansatz}
\ee
this leads to the finite-volume estimator for the interface tension
\be
a^3\hat{\sigma}=\ltz(c_1+c_2)\,.
\label{sigmaest}
\ee
The inverse-linear term in eq.~(\ref{fitansatz}) is motivated by the fact
that for $\Delta\kappa\equiv|\kappa_i-\kappa_c|\ll1$ the free energy change
between the two phases behaves as $\Delta F\simeq\cl{O}(\Delta\kappa)$.
The probability $p$ for an interface being at $\z_0>\z$ is essentially
given by $\exp\{-\mbox{const}\cdot\Delta\kappa(\z-\z_0)\}$ and therefore
$\int\dt\z\,p\simeq 1/\Delta\kappa$.

\begin{figure}[h]
\begin{center}
\epsfig{file=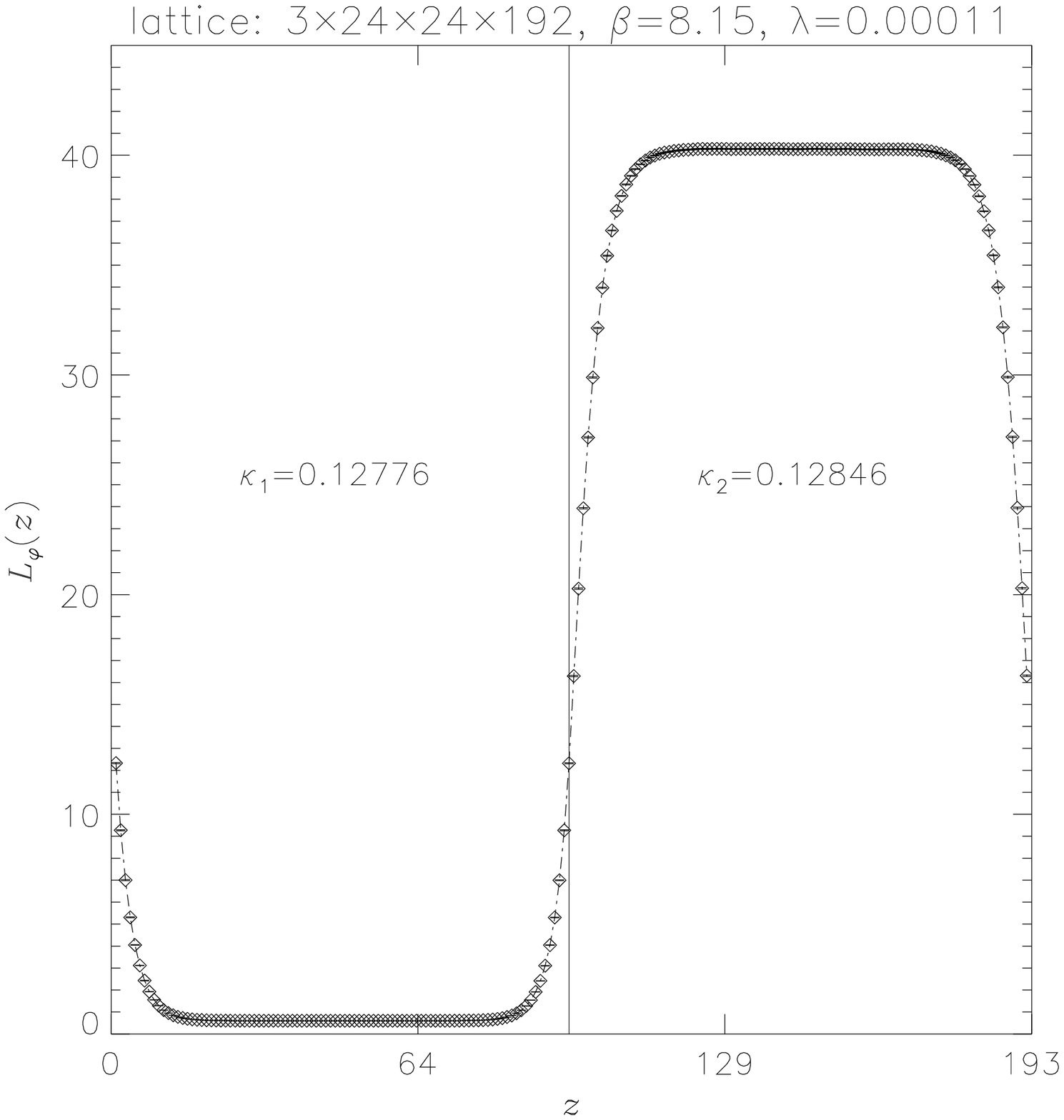,width=8cm}
\epsfig{file=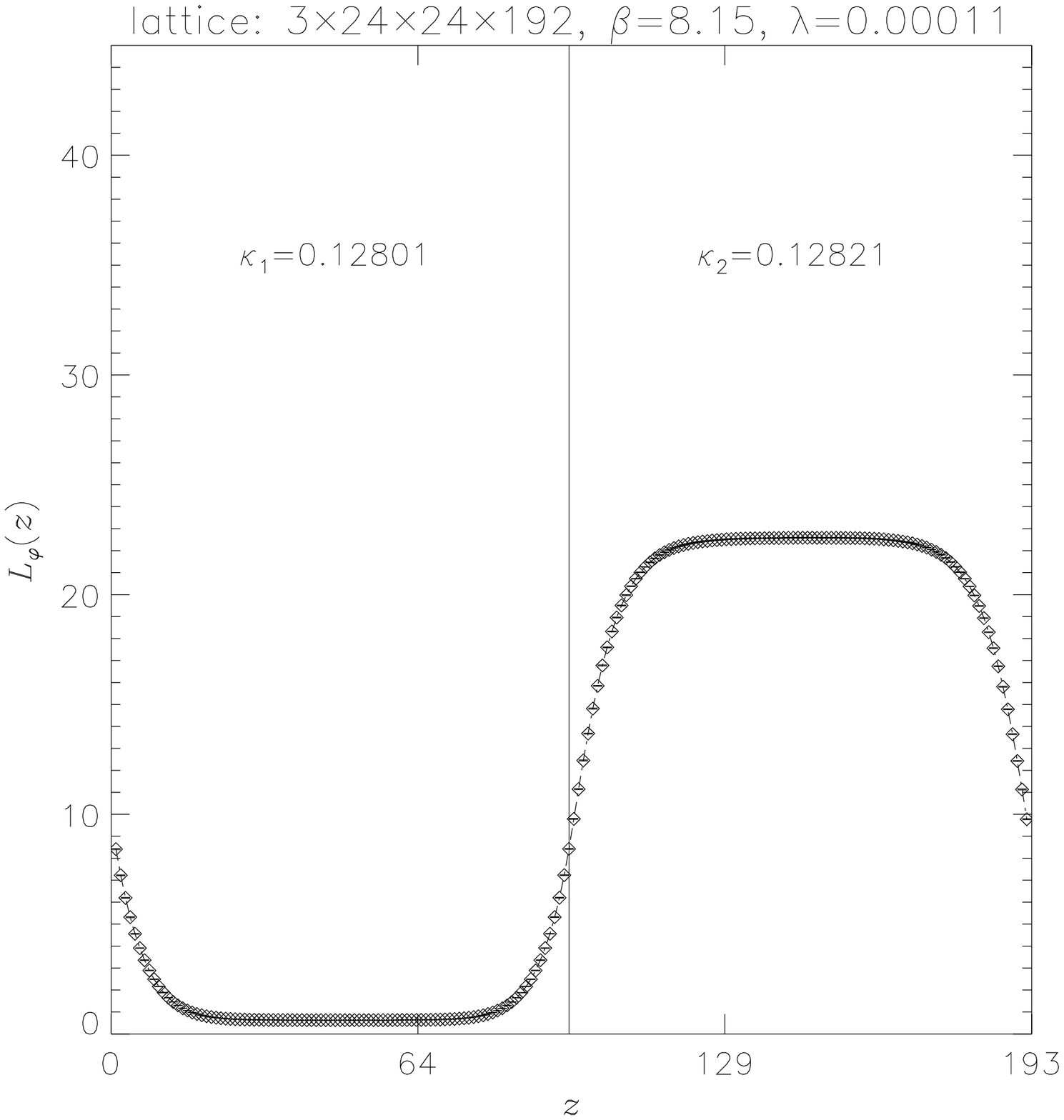,width=8cm}
\parbox{16cm}{
\caption{\label{LphiPlot} \sl Two-phase profiles of the z--slice expectation
                              value $L_{\varphi}(\z)$ of $L_{\varphi}$ for
                              our largest and smallest $\kappa$--intervals.}
}
\end{center}
\end{figure}

Thus one proceeds in a similar way as for the $\kappa_c$--determination,
but due to practical limitations on $\ltz$, one has to prevent the 
interfaces from touching each other by a choice of large enough 
$\kappa$--intervals.
Figure~\ref{LphiPlot} displays typical two-phase structures from the
2--$\kappa$ method.
Note that the plateaus become narrower for smaller $\kappa$--intervals,
especially in the broken phase, but the phases are still clearly developed,
and the interfaces continue to exist.
%
%%% Beginn Figur %%%
%%% Ende Figur %%%
%

The results of our simulations are presented in table~\ref{ResLphi}.
Owing to correlations between $L_{\varphi}^{(1)}$ and $L_{\varphi}^{(2)}$,
the statistical errors on $\Delta L_{\varphi}$ are usually smaller
than those of $L_{\varphi}^{(2)}$.
This was already suggested by the autocorrelations in figure~\ref{AcfPlot}
and is caused by shifts of the interfaces during the simulation, which 
cancel out in $\Delta L_{\varphi}$ to some extent.
%
%%% Beginn Tabelle %%%
\begin{table}[h]
\begin{center}
\begin{tabular}{|c|c|c||c|c|c|}
\hline
  $\kappa_1$ & $\kappa_2$ & sweeps & $L_{\varphi}^{(1)}$
& $L_{\varphi}^{(2)}$ & $\Delta L_{\varphi}$ \\
\hline\hline
  0.12776 & 0.12846 &  5000 & 1.5161(24) & 37.8216(70) & 36.3055(67) \\
  0.12781 & 0.12841 &  5000 & 1.5149(26) & 34.7287(68) & 33.2138(60) \\
  0.12786 & 0.12836 &  5000 & 1.5285(29) & 31.5085(77) & 29.9799(76) \\
  0.12791 & 0.12831 & 10000 & 1.5398(19) & 28.1144(55) & 26.5746(57) \\
  0.12796 & 0.12826 & 10000 & 1.5616(30) & 24.5264(76) & 22.9648(71) \\
  0.12801 & 0.12821 & 20000 & 1.6019(32) & 20.5877(66) & 18.9858(59) \\
\hline
\end{tabular}
\parbox{12cm}{
\caption{\label{ResLphi} \sl Results for $L_\varphi^{(1)}$,
                             $L_\varphi^{(2)}$ and $\Delta L_\varphi$ on a
                             $3\times24^2\times192$ lattice.
                             The errors are obtained by binning.}
}
\end{center}
\end{table}
%%% Ende Tabelle %%%
%

In order to give a reliable estimate for the statistical error when fitting
these $\varphi$--link averages to the functions in eq.~(\ref{fitansatz}),
we take their correlations into account by a \emph{bootstrap analy\-sis}
\cite{E79GGKPSW87}.
The characteristic ingredient is to calculate secondary quantities from
bootstrap subsamples randomly taken with repetition from the original
measurements, which itself can be interpreted as the empirical probability
distribution of the observables under consideration.
The errors of the fit parameters are extracted as half of the central
68.3\%--interval of their distributions from fits performed on these sample
averages.
A more extensive description of this method in the same context is contained
in ref.~\cite{CFHH95}.

We quote our final result for two four-parameter fits with $\chi^2$--values of
2.58 and 0.82.
Using (\ref{sigmaest}), we find
\be
\left(\frac{\hat{\sigma}}{T_c^3}\right)_{L_{\varphi}}=0.764(52+47)\,,
\label{sigma4fit}
\ee
whereby $aT_c=1/L_t=\frac{1}{3}$ in lattice units.
The error consists of two parts, coming from a bootstrap analysis with
10000 iterations and the uncertainty of $\kappa_c$ in
eq.~(\ref{kappares}).
%
%%% Beginn Figur %%%
\begin{figure}[h]
\begin{center}
\epsfig{file=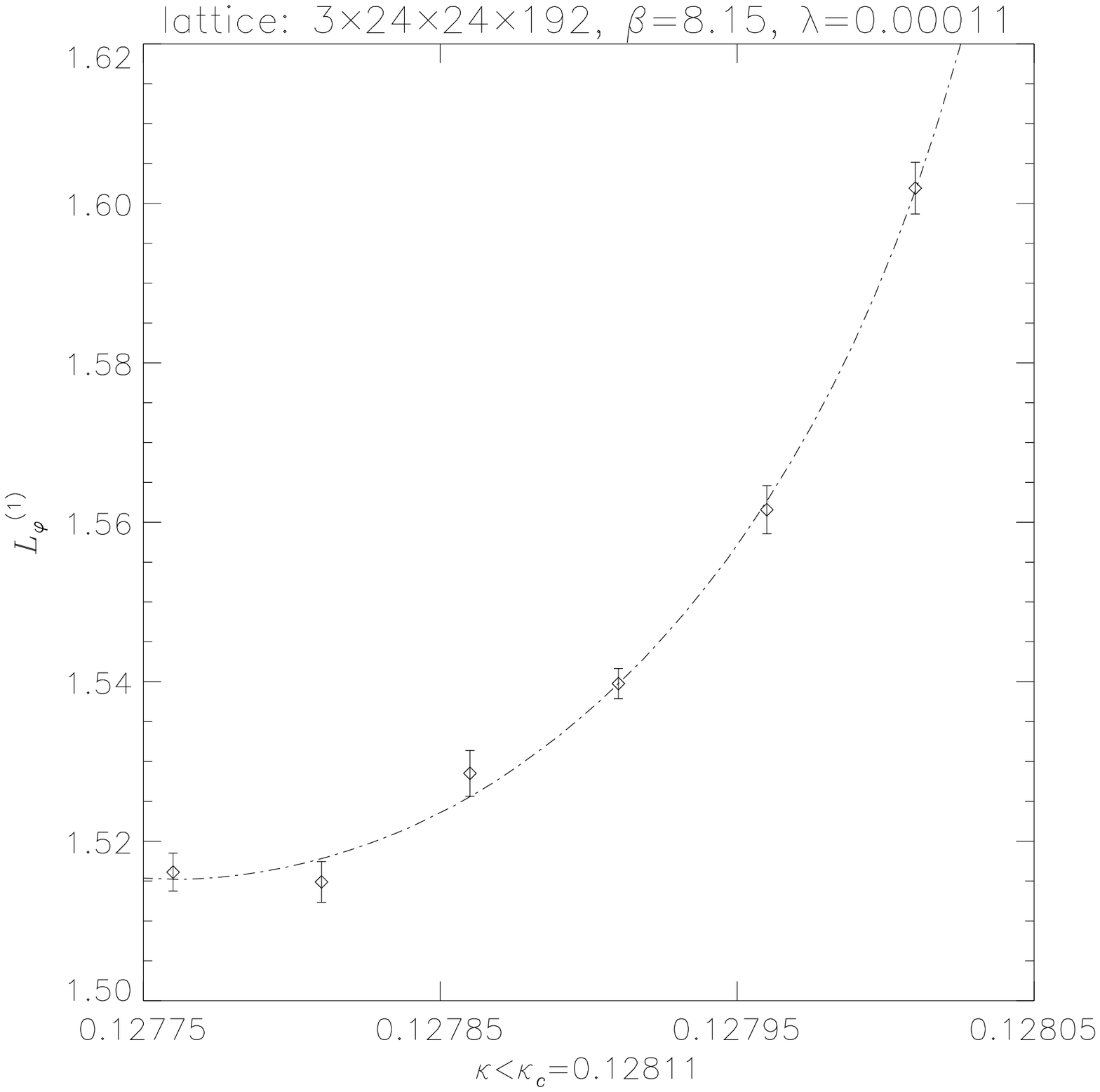,width=8cm}
\epsfig{file=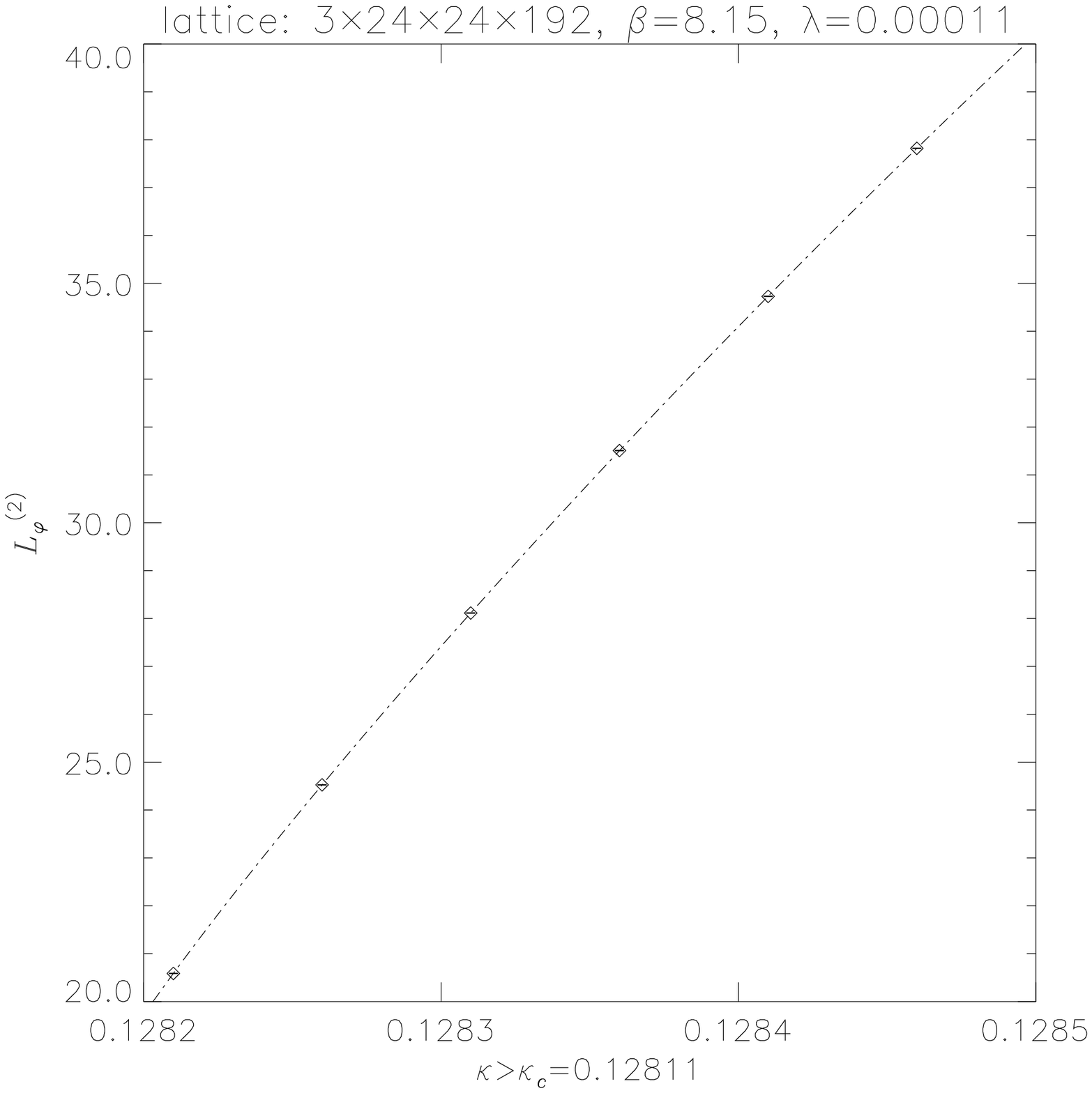,width=8cm}
\parbox{16cm}{
\caption{\label{LfitPlot} \sl Four-parameter least-squares fits of
                              $L_{\varphi}^{(i)}$, $i=1,2$, separately
                              in each phase.}
}
\end{center}
\end{figure}
%%% Ende Figur %%%
%

At this place some explanations about the number of necessary fit
parameters are in order.
In the examinations of $L_t=2$ lattices \cite{FHJJM95,CFHH95} a
three-parameter fit was sufficient to give a reliable value for the 
interface tension; an inclusion of higher order terms gave no improvement.
The situation for $L_t=3$ seems to be different for the following reason:
Basically, one has to keep in mind that the expansion in
eq.~(\ref{fitansatz}) is only a phenomenological ansatz.
Furthermore the lattice volume $\Omega$ has increased, and the phase
transition, which in a strict sense is only present in infinite volumes,
is more pronounced.
Consequently, the contribution of the higher Laurent coefficients becomes
more important for the slopes at larger $\kappa$--intervals.

In fact, the three-parameter fit has no satisfactory $\chi^2$.
A five-parameter fit gives \mbox{
$\hat{\sigma}/T_c^3=0.85(20+5)$
}
and $\chi^2$--values equal to 1.42 and 0.82.
This is compatible to (\ref{sigma4fit}) within errors, but fairly sensitive
to the number of fitted data points.
A more careful inspection of the bootstrap calculations reveals that the
last fit parameter is not very significant in such cases, where its
bootstrap error is roughly as large as its value; this holds true for
the five-parameter fits and also the four-parameter fit in the symmetric 
phase.
So for the sake of completeness we combined the three-parameter fit in the
symmetric phase with the four-parameter fit in the broken phase to
$\hat{\sigma}/T_c^3=0.793$, although we are nevertheless convinced of the
four-parameter fit in both phases to lead to the most reasonable result.

\begin{figure}[h]
\begin{center}
\epsfig{file=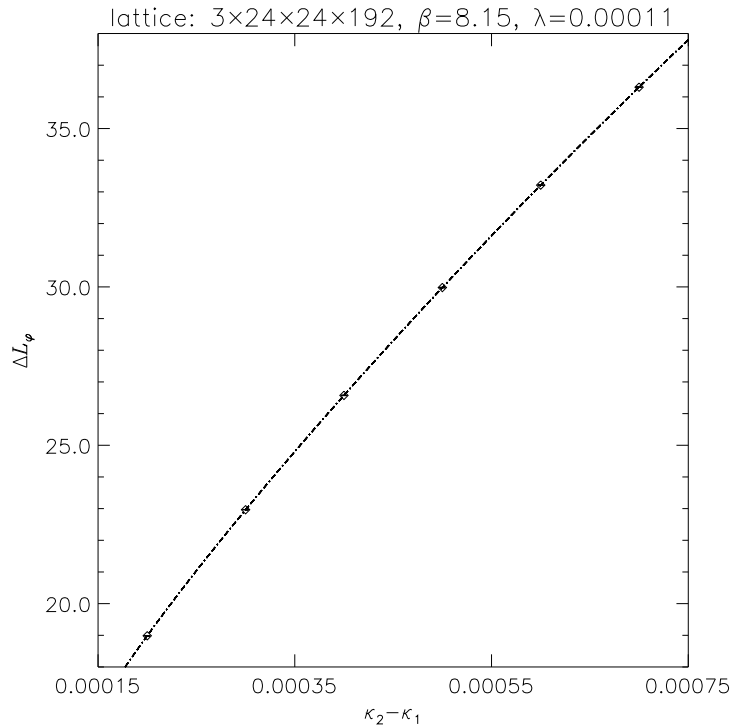,width=10cm}
\parbox{10cm}{
\caption{\label{DLfitPlot} \sl Four-parameter least-squares fit of
                               $\Delta L_{\varphi}$ as a function of
                               $\kappa_2-\kappa_1$.}
}
\end{center}
\end{figure}

The preceding remarks on the $\varphi$--link correlations should have
made clear that $\Delta L_{\varphi}$ is the most natural variable for
estimating the interface tension via eq.~(\ref{sigmares}).
This requires that the chosen $\kappa$--intervals are symmetric with respect
to $\kappa_c$.
A four-parameter fit to an ansatz similar to (\ref{fitansatz}) for
$\Delta L_{\varphi}$ yields
\be
\left(\frac{\hat{\sigma}}{T_c^3}\right)_{\Delta L_{\varphi}}=0.767(53)
\label{sigma4Dfit}
\ee
with $\chi^2=0.39$ and is illustrated in figure~\ref{DLfitPlot}.
Here we only quote the statistical error, which now comes from 1000 normally
distributed random data, since the different
$\Delta L_{\varphi}$--averages are uncorrelated.
Note the perfect agreement of this result, and especially of its error,
with the numbers in eq.~(\ref{sigma4fit}) above.
%
%%% Beginn Figur %%%
%%% Ende Figur %%%
%
\section{Discussion}
We have determined the interface tension of the four-dimensional 
SU(2)--Higgs model with the two-coupling method.
All estimates from acceptable fits with a reasonable number of parameters
show a very good consistency.

To our knowledge this is the first attempt to supply information about the
scaling of $\sigma$ when going over to a finer lattice.
Compared to the $L_t=2$ results
\be
\left(\frac{\hat{\sigma}}{T_c^3}\right)_{
      \scriptsize \mbox{2--$\kappa$}}=0.84(16)\,,\quad
\left(\frac{\sigma}{T_c^3}\right)_{
      \scriptsize \mbox{hist}}=0.83(4)
\label{l2sigma}
\ee
from refs.~\cite{CFHJJM9495,FHJJM95} --- the second number referring to the 
histogram method --- the accuracy of the 2--$\kappa$ estimate has been 
improved to the value \mbox{
$\hat{\sigma}/T_c^3=0.76(10)$
}
in eq.~(\ref{sigma4fit}).
The observed small deviation between $L_t=2$ and $L_t=3$ confirms the 
smallness of scaling violations, as e.g.~also found recently for the critical
temperature in \cite{CFHJM96}.

Finally, we confront our result with the perturbative estimate 
\cite{FH94BFH9495} up to order $g^4$,$\lambda_0^2$
\be
\left(\frac{\sigma}{T_c^3}\right)_{\scriptsize \mbox{pert}}=0.78(1)\,,
\label{sigmapert}
\ee
with an error coming from the uncertainties in the renormalized parameters
on the lattice.
As expected, the agreement with perturbation theory on a quantitative level
in this range of the Higgs boson mass is excellent.
\subsection*{Acknowledgements}
We thank Z.~Fodor, J.~Ignatius, K.~Jansen, I.~Montvay and G.~M\"unster for 
some helpful comments.

All numerical simulations have been done on the CRAY Y-MP8/864 of HLRZ in
J\"ulich, Germany, which offers 64--bit floating point precision.
%
%%%%%%%%%%%%%%%%%%%%%%%%%%%%%%%%%%%%%%%%%%%%%%%%%%%%%%%%%%%%%%%%%%%%%%%%%%%%
%
% \newpage
%

%
\end{document}